\documentclass[12pt]{iopart}

\usepackage{amssymb}
\usepackage{graphicx}
\usepackage{color}
\usepackage{hyperref}
\usepackage{lineno}

\usepackage{ulem}

\begin{document}

%\linenumbers

\title{Amorphous optical coatings of present gravitational-wave interferometers}

\date{\today}

\author{M Granata$ˆ1$, A Amato$ˆ1$, L. Balzarini$ˆ1$\footnote{Deceased, August 2018.},M Canepa$ˆ{2,3}$, J Degallaix$ˆ1$, D Forest$ˆ1$, V Dolique$ˆ1$\footnote{Current affiliation: Universit\'e de Lyon, ENS de Lyon, Universit\'e Claude Bernard Lyon 1, CNRS, Laboratoire de Physique, F-69342 Lyon, France}, L Mereni$ˆ1$, C Michel$ˆ1$, L Pinard$ˆ1$, B Sassolas$ˆ1$, J Teillon$ˆ1$ and G Cagnoli$ˆ{4}$}
\address{$ˆ1$ Laboratoire des Mat\'eriaux Avanc\'es-IP2I, CNRS, Universit\'e de Lyon, F-69622 Villeurbanne, France}
\address{$ˆ2$ OPTMATLAB, Dipartimento di Fisica, Universit\`a di Genova, Via Dodecaneso 33, 16146 Genova, Italy}
\address{$ˆ3$ INFN, Sezione di Genova, Via Dodecaneso 33, 16146 Genova, Italy}
\address{$ˆ4$ Universit\'e de Lyon, Universit\'e Claude Bernard Lyon 1, CNRS, Institut Lumi\`ere Mati\`ere, F-69622, Villeurbanne, France}

\ead{m.granata@lma.in2p3.fr}

\begin{abstract}
We report on the results of an extensive campaign of optical and mechanical characterization of the ion-beam sputtered oxide layers (Ta$_2$O$_5$, TiO$_2$, Ta$_2$O$_5$-TiO$_2$, SiO$_2$) within the high-reflection coatings of the Advanced LIGO, Advanced Virgo and KAGRA gravitational-wave detectors: refractive index, thickness, optical absorption, composition, density, internal friction and elastic constants have been measured; the impact of deposition rate and post-deposition annealing on coating internal friction has been assessed. For Ta$_2$O$_5$ and SiO$_2$ layers, coating internal friction increases with the deposition rate, whereas the annealing treatment either erases or largely reduces the gap between samples with different deposition history. For Ta$_2$O$_5$-TiO$_2$ layers, the reduction of internal friction due to TiO$_2$ doping becomes effective only if coupled with annealing. All measured samples showed a weak dependence of internal friction on frequency ($\phi_c(f) = af^{b}$, with $-0.208 < b < 0.140$ depending on the coating material considered). SiO$_2$ films showed a mode-dependent loss branching, likely due to spurious losses at the coated edge of the samples. The reference loss values of the Advanced LIGO and Advanced Virgo input (ITM) and end (ETM) mirror HR coatings have been updated by using our estimated value of Young's modulus of Ta$_2$O$_5$-TiO$_2$ layers (120 GPa) and are about 10\% higher than previous estimations.\\

\noindent{\it Thin films; internal friction; thermal noise; gravitational-wave detectors.}
\end{abstract}

\maketitle

\section{Introduction}
On September 14th 2015 at 09:50:45 UTC, a century after the fundamental predictions of Einstein \cite{Einstein16, Einstein18}, the two detectors of the Advanced Laser Interferometer Gravitational-Wave Observatory (Advanced LIGO) \cite{aLIGO} simultaneously observed a transient gravitational-wave signal from the merger of two stellar-mass black holes \cite{GW150914}. This event was the first direct detection of gravitational waves and the first observation of a binary black hole merger, and marked the beginning of gravitational astronomy. Since then other detections followed \cite{GW151226, GW170104, GW170608, GW170814, GWcatalog}, including a multi-messenger binary neutron-star merger \cite{GW170817,GW170817multim}.

The Laboratoire des Mat\'{e}riaux Avanc\'{e}s (LMA) has provided the high-reflection (HR) and anti-reflective (AR) coatings for the core optics of the Advanced LIGO, Advanced Virgo \cite{AdVirgo} and KAGRA \cite{KAGRA} gravitational-wave interferometers \cite{Pinard17}. In such detectors, those large and massive suspended mirrors (up to $\varnothing = 35$ cm, $t = 20$ cm and $m = 40$ kg) play the crucial role of gravitational-field probes for the astrophysical signals \cite{Abbott16det}.

The HR coatings are Bragg reflectors of alternate layers of ion-beam-sputtered (IBS) low- and high-refractive-index materials. Initially, tantalum pentoxide (Ta$_2$O$_5$, also known as {\it tantala}) and silicon dioxide (SiO$_2$, {\it silica}) had been chosen as high- and low-index materials, respectively, because of their low optical absorption at the typical wavelength of operation of the detectors ($\lambda_0=1064$ nm) \cite{Beauville04}. Further development of the tantala layers was driven by thermal-noise issues.

In gravitational-wave interferometers, thermal noise arises from fluctuations of the mirror surface under the random motion of particles in coatings and substrates \cite{Saulson90, Levin98}. Its intensity is determined by the amount of internal friction within the mirror materials, via the fluctuation-dissipation theorem \cite{Callen52}: the higher the mechanical-energy loss, the higher the thermal noise level. As the coating loss is usually several orders of magnitude larger than that of the substrate \cite{Crooks02, Harry02}, in the last two decades a considerable research effort has been committed to the investigation and the reduction of thermal noise in optical coatings. 

Within the stack, tantala proved to be substantially more dissipative than silica \cite{Penn03, Crooks04}, making it the dominant source of coating loss. This loss was remarkably decreased by applying a titanium dioxide (TiO$_2$, {\it titania}) doping to tantala, a procedure developed by the LMA \cite{Comtet07} for the LIGO Scientific Collaboration \cite{Harry06, Harry07}. Eventually, titania doping proved to be beneficial to the optical absorption of the HR coating as well, reducing it \cite{Flaminio10}. In the meantime, a technique to optimize the coating design in order to dilute the loss contribution of the high-index material had also been developed \cite{Agresti06}, further decreasing the resulting coating thermal noise \cite{Villar10}.

In this paper, we report on the results of an extensive campaign of optical and mechanical characterization of the materials within the HR coatings of all the present kilometer-scale gravitational-wave interferometers. The relevance of these results is threefold: i) we provide several parameters which are required to predict the coating thermal noise in the detectors; ii) we point out the relevance of the synthesis process for the coatings' properties, in particular with respect to mechanical loss; iii) we set reference values for further research and development of low-noise coatings.
%
% ---------------------------------------------
%
\section{Experiment}
\label{SECexp}
The amorphous IBS coating materials studied for this work are silica, tantala, titania and titania-doped tantala. They have been deposited with three different coaters at LMA: the custom-developed so-called {\it DIBS} and {\it Grand Coater} (GC) and a commercially available Veeco SPECTOR. The GC is used to coat the mirrors of gravitational-wave detectors.

The coatings have been deposited on different kind of substrates for different purposes: fused-silica witness samples ($\varnothing = 1$", $t = 6$ mm) and silicon wafers ($\varnothing = 3$", $t = 1$ mm) for the optical characterization, fused-silica disk-shaped resonators ($\varnothing = $ 50 or 75 mm, $t = 1$ mm, with two parallel flats) for the mechanical characterization. Prior to coating deposition, to release the internal stress due to manufacturing and to induce relaxation, the fused-silica disks have been annealed in air at 900 $^\circ$C for 10 hours.

As part of the standard post-deposition process adopted for the production of gravitational-wave detectors' mirrors, in order to decrease both the internal stress and the optical absorption of the coatings, all the samples have been annealed in air at 500 $^\circ$C for 10 hours. This post-deposition annealing eventually turns out to be beneficial also for thermal noise, since it significantly decreases the coating loss.
%
% -------------------------
%
\subsection{Optical characterization}
We measured the film refractive index and thickness, by transmission spectrophotometry through coated fused-silica witness samples and by reflection spectroscopic ellipsometry on coated silicon wafers.

Spectrophotometric measurements have been carried out with a Perkin Elmer Lambda 1050 spectrophotometer. Spectra have been acquired at normal incidence in the 400-1400 nm range. Film refractive index and thickness were first evaluated using the envelope method \cite{Cisneros98}, then these results were used as initial values in a numerical least-square regression analysis. In the model, the adjustable parameters were the thickness and the ($B_i$, $C_i$) coefficients of the Sellmeier dispersion equation:
\begin{equation}
n^2 = 1 + \sum_{i=1}^{3} \frac{B_i \lambda^2}{\lambda^2 - C_i}\ .
\end{equation}

We used used two J.A. Woollam Co. instruments for the ellipsometric analysis, a VASE for the 190-1100 nm range and a M-2000 for the 245-1680 nm range. The wide wavelength range swept with both ellipsometers allowed us to extend the analysis from ultraviolet to infrared (0.7 - 6.5 eV). The optical response of the substrates has been characterized with prior dedicated measurements. To maximize the response of the instruments, coating spectra have been acquired for three different angles of incidence of the light ($\theta = 50^\circ$, $55^\circ$, $60^\circ$), chosen to be close to the Brewster angle of each coating material ($\theta_B \sim 55^\circ$ for silica and $\sim 64^\circ$ for tantala, for instance). The refractive index and thickness of the films have been derived by comparing the experimental data with simulations based on realistic optical models \cite{Fujiwara07}. More details about our ellipsometric analysis can be found in a dedicated article \cite{Amato19}.

Finally, we measured the coating optical absorption at $\lambda_0=1064$ nm through photo-thermal deflection \cite{Boccara80}.% The details of our setup are available elsewhere \cite{Degallaix13}.
%
% -------------------------
%
\subsection{Mechanical characterization}
To measure the coating mass, we have used an analytical balance and measured the mass of the disks before and after the deposition, as well as after the annealing. To estimate the surface area coated in the deposition process, we have measured the diameter and the flat spacing of the disks with a Vernier caliper. As the coating thickness is known with high accuracy from the optical characterization, the coating density $\rho_c$ could be straightforwardly estimated as the mass-to-volume ratio.

To measure the coating loss, we have applied the ring-down method \cite{Nowick72} to the disks and measured the ring-down time of their vibrational modes. In each sample, for the $k$-th mode of frequency $f_k$ and ring-down time $\tau_k$, the measured loss is $\phi_k = (\pi f_k \tau_k)^{-1}$. The coating loss $\phi^c_k$ can be written
\begin{equation}
\label{EQcoatLossKth}
\phi^c_k = \left[\ \phi_k + (D_k-1)\phi_k^s\ \right]/D_k \ ,
\end{equation}
where $\phi^s_k$ is the measured loss of the bare substrate. $D_k$ is the \textit{dilution factor}, defined as the ratio of the elastic energy of the coating, $E_c$, to the elastic energy of the coated disk, $E = E_c + E_s$, where $E_s$ is the elastic energy of the substrate. This ratio actually depends on the mode shape, which in turn is determined by the number of $r$ radial and $a$ azimuthal nodes \cite{McMahon64}, $r$ and $a$ respectively, denoted by the pair $(r, a)_k$. $D_k$ is thus mode-dependent and can be written as a function of the frequencies $f^s_k$, $f_k$ and of the masses $m^s$, $m$ of the sample before and after the coating deposition, respectively \cite{Li14}:
\begin{equation}
\label{EQdilFact}
D_k = 1 -  \left( \frac{f^s_k}{f_k} \right)^2 \frac{m^s}{m} \ .
\end{equation}

We have used a Gentle Nodal Suspension (GeNS) \cite{Cesarini09} to suspend the disks from the center, in order to avoid systematic damping from suspension. The system was placed inside a vacuum enclosure at $p \leq 10^{-6}$ mbar to prevent residual-gas damping. We have measured up to sixteen modes on each disk, sampling the coating loss in the 1-30 kHz band. This sampling partially overlaps with the detection band of ground-based gravitational-wave interferometers (10-10$^4$ Hz).

For each coated disk, we have estimated the Young's modulus $Y_c$ and the Poisson's ratio $\nu_c$ of the coating materials by adjusting finite-element simulations to match the measured dilution factors from Eq.(\ref{EQdilFact}): we found the set of values ($Y_c$, $\nu_c$) minimizing the least-square figure of merit
\begin{equation}
\label{EQmerFunD}
m_D = \sum_{k} \left[ \frac{D_k^{\textrm{meas}}-D_k^{\textrm{sim}}}{\sigma_k^{\textrm{meas}}} \right]^2 \ ,
\end{equation}
where $D_k^{\textrm{sim}}$ and ($D_k^{\textrm{meas}} \pm \sigma_k^{\textrm{meas}}$) are the simulated and measured dilution factors, respectively ($\sigma_k^{\textrm{meas}}$ is the measurement uncertainty). In this method, knowledge of the substrate parameters is critical: dimensions have been assigned measured values, values of density ($\rho = 2202$ g/cm$^3$), Young's modulus ($Y = 73.2$ GPa) and Poisson's ratio ($\nu = 0.17$) have been taken from the literature \cite{McSkimin53}; in a dedicated subset of simulations of the bare substrates, thickness $t$ has been adjusted to minimize a merit function $m_t$ of the same form as Eq.(\ref{EQmerFunD}) for simulated and measured mode frequencies ($f_k^{\textrm{sim}}$ and $f_k^{\textrm{meas}}$, respectively). For a few disks, we independently measured $t$ with a micrometer and found that the discrepancy with the fitted values is less than 2\%.
\begin{figure} 
\centering
	\includegraphics{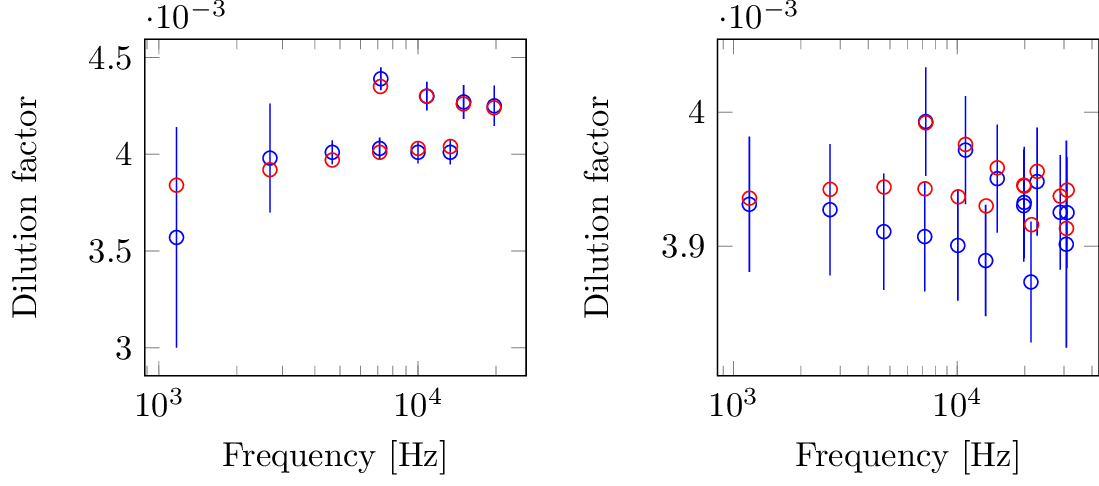}
	\caption{(Color online) Comparison between measured (blue) and simulated (red) dilution factors, for Ta$_2$O$_5$ (left) and SiO$_2$ (right) coatings annealed in air at 500 $^\circ$C during 10 hours.}
	\label{FIGdilFact}
\end{figure}
Fig. \ref{FIGdilFact} shows the results of the overall process, comparing measured and simulated dilution factors of representative samples of tantala and silica. Dilution factors of modes with a different number of radial nodes lie on distinct curves, which we have called {\it mode families}; thus, for example, modes with $(0, a)_k$ and modes with $(1, a)_k$ belong to two different families (called {\it butterfly} and {\it mixed} modes, respectively). For a given substrate with ($Y$, $\nu$), dilution factors are determined by the coating elastic constants: their average value depends on the $Y_c/Y$ ratio, as in the case of flexural modes of cantilevers \cite{Heptonstall06}, whereas the separation between mode families increases as the difference $\vert \nu_c - \nu \vert$ grows.

Further details about our GeNS system and finite-element simulations are available elsewhere \cite{Granata16}.
%
% ---------------------------------------------
%
\section{Results}
\label{SECres}
\begin{figure} 
\centering
	\includegraphics{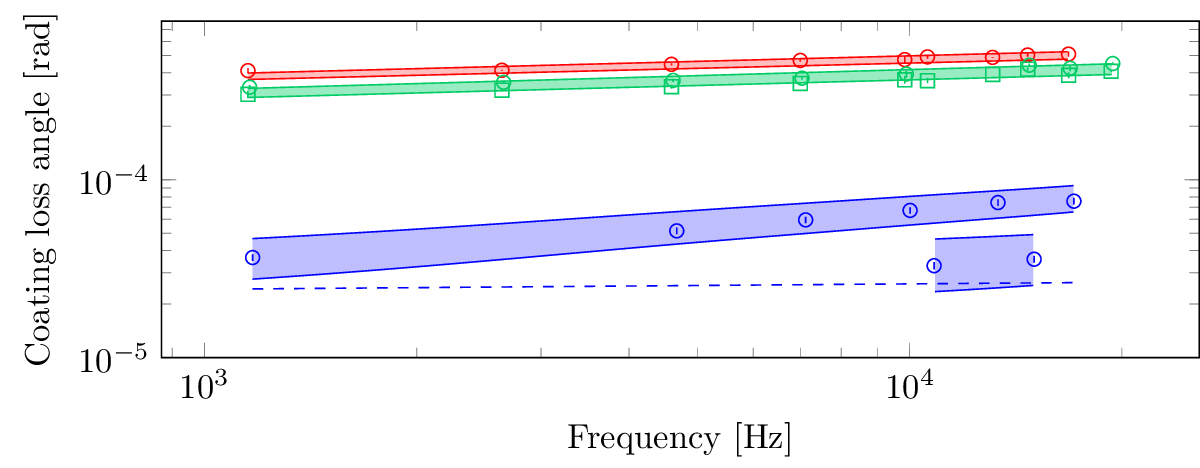}
	\caption{Mechanical loss of coating materials of Advanced LIGO, Advanced Virgo and KAGRA after annealing: Ta$_2$O$_5$ (red),  Ta$_2$O$_5$-TiO$_2$ (green), SiO$_2$ (blue). Different markers denote distinct samples; error bars are shown, though barely visible. Shaded regions represent uncertainties from fitting a frequency-dependent loss model to each data set: $\phi_c(f) = af^{b}$ via least-squares linear regression for Ta$_2$O$_5$ and Ta$_2$O$_5$-TiO$_2$, $\phi_c(f) = af^{b} + \epsilon d\phi_e$ via iterative non-linear regression for SiO$_2$; the dashed curve shows the behavior of the $af^{b}$ term only for SiO$_2$.}
	\label{FIGcoatLossGC}
\end{figure}

In order to stress the impact of deposition parameters and post-deposition annealing, each coating material is discussed separately in the following. Also, because of their relevance, results of coatings materials of gravitational-wave interferometers are summarized separately in Tables \ref{TABLEoptParGC} and \ref{TABLEmechParGC} and in Fig. \ref{FIGcoatLossGC}.

Values of density, Young's modulus and Poisson's ratio are given with 1$\sigma$ uncertainty. To characterize the loss behavior, we have fitted a frequency-dependent model $\phi_c(f) = af^{b}$ to the data of tantala, titania and titania-doped tantala coatings by linear regression; for silica coatings, in order to account for the observed loss branching, a term $\epsilon d\phi_e$ had to be included in the model ($\phi_c(f) = af^{b} + \epsilon d\phi_e$) and numerical non-linear regression has been used. This term quantifies the amount of spurious loss $\phi_e$ in a thin surface layer of the coated disk edge \cite{Cagnoli18}.

For conciseness, the discrepancy between the results of spectrophotometry and ellispometry being less than 3\%, only ellipsometric values are reported (unless otherwise explicitly mentioned). Refractive index is given for $\lambda_0=1064$ nm, as well as for the alternative wavelength $\lambda=1550$ nm of future detectors such as the Einstein Telescope \cite{Hild11, Abernathy11} and Cosmic Explorer \cite{Abbott17}. Upon annealing, the optical absorption of all the coatings (except for the crystallized TiO$_2$) decreased to sub-ppm (parts per million) values, which corresponds to an extinction coefficient of $10^{-7} < k < 10^{-6}$.
\begin{table}
\caption{\label{TABLEoptParGC}Refractive index of coating materials of Advanced LIGO, Advanced Virgo and KAGRA.}
	\begin{indented}
	\lineup	
	\item[] \begin{tabular}{@{}lcc}
		\br
			 				& 1064 nm 		& 1550 nm\\
		\mr
	 	Ta$_2$O$_5$ 		& 2.05 $\pm$ 0.01	& 2.03 $\pm$ 0.01\\
	 	Ta$_2$O$_5$-TiO$_2$ & 2.09 $\pm$ 0.01	& 2.08 $\pm$ 0.01\\
	 	SiO$_2$				& 1.45 $\pm$ 0.01	& 1.45 $\pm$ 0.01\\
	\br
			\end{tabular}
	\end{indented}
\end{table}
% ---
\begin{table}
\caption{\label{TABLEmechParGC}Mechanical parameters of coating materials of Advanced LIGO, Advanced Virgo and KAGRA.}
\begin{indented}
\lineup	
\item[] \begin{tabular}{lccccc}
	\br 		
	 					& $\rho_c$ [g/cm$^3$]	& a [10$^{-4}$ rad Hz$^{-b}$]	& b						& $Y_c$ [GPa]	& $\nu_c$\\
	\mr
	Ta$_2$O$_5$ 		& 7.33 $\pm$ 0.06		& 1.88 $\pm$ 0.06				& 0.101 $\pm$ 0.004		& 117 $\pm$ 1	& 0.28 $\pm$ 0.01\\
 	Ta$_2$O$_5$-TiO$_2$ & 6.65 $\pm$ 0.07		& 1.43 $\pm$ 0.07				& 0.109 $\pm$ 0.005		& 120 $\pm$ 4	& 0.29 $\pm$ 0.01\\
 	SiO$_2$ 			& 2.20 $\pm$ 0.04		& 0.20 $\pm$ 0.04				& 0.030 $\pm$ 0.024		& 70 $\pm$ 1	& 0.19 $\pm$ 0.01\\
	\br
\end{tabular}
\end{indented}	
\end{table}
% ---
%
% -------------------------
%
\subsection{Ta$_2$O$_5$}
Before annealing, the tantala films had similar refractive index and equal Young's modulus and Poisson's ratio, whereas their density and loss were different.

Interestingly, the loss seems to be related to the coating deposition rate, determined by the energy and the flux of particles of the sputtering beam and by the configuration of the coating chamber. In our case, the GC provided the slowest rate and lowest loss, whereas the SPECTOR the fastest rate (3 \AA/s) and the highest loss; thus, according to our results, the faster the deposition rate, the higher the loss.

The situation changed radically after the annealing, when all the films exhibited equal and significantly lower loss, as if their deposition history had been completely erased. This outcome seems to suggest that 500 $^\circ$C in-air annealing for 10 hours brings the structure of tantala coatings down to a stable optimal configuration for lowest loss, in agreement with observations that higher annealing temperatures or longer duration do not decrease loss further \cite{Amato18}. The same {\it erasing} effect was observed later in an independent experiment, where sputtered tantala coatings (IBS, magnetron) had been annealed after being deposited on heated substrates \cite{Vajente18}. Finally, refractive index and density also slightly decreased after annealing, as the physical thickness of all the films increased.

By now, there have been many experimental studies on the internal friction of tantala coatings \cite{Penn03, Comtet07, Flaminio10, Villar10, Li14, Martin10, Principe15, Cesarini11}. However, the comparison with our results is not straightforward: previous works used coatings deposited with different conditions \cite{Villar10, Martin10} and/or stacked with other layers \cite{Penn03, Villar10, Principe15}; moreover, their analyses relied on assumptions made on the value of the coating Young's modulus \cite{Penn03, Comtet07, Flaminio10, Li14, Martin10, Principe15, Cesarini11}. Probably, the fairest comparison would be with annealed tantala coatings deposited with the GC on cantilever blades \cite{Comtet07} under identical conditions, featuring a constant loss of \mbox{(3.0 $\pm$ 0.1)$\cdot 10^{-4}$ rad} in the \mbox{60-1100 Hz} band; by extrapolating our results of the annealed samples to lower frequencies, we obtain \mbox{(2.8 $\pm$ 0.1)$\cdot 10^{-4}$ rad} at 60 Hz and \mbox{(3.8 $\pm$ 0.2)$\cdot 10^{-4}$ rad} at 1100 Hz, matching the value from cantilever blades at 110 Hz. By correcting the cantilever blades results by our value of coating Young's modulus we would obtain 3.9 $\cdot$ 10$^{-4}$ rad, so that the match with our results would be shifted to 1260 Hz. At 8.9 kHz, our results also match the value of (4.7 $\pm$ 0.6)$\cdot 10^{-4}$ rad from quadrature-phase interferometry \cite{Li14} on our SPECTOR films, obtained assuming constant loss but with no further assumptions on the coating Young's modulus; as a matter of fact, those results already pointed to an actual Young's modulus of 118 GPa, very close to our value (121 $\pm$ 2 GPa).

If measured via nano-indentation, the Young's modulus of IBS tantala coatings appears to be about 140 GPa \cite{Cetinorgu09, Abernathy14}, i.e. about 18\% higher than our value. This difference could be explained by the nature of the films, deposited with different conditions, and by the fact that results from nano-indentation are model dependent and rely on assumptions about the coating Poisson's ratio. Furthermore, nano-indentations of the same coating deposited on different substrates might give different results: our tantala SPECTOR coatings yielded a reduced coating Young's modulus of 130 $\pm$ 3 GPa on silica witness samples and of 100 $\pm$ 3 GPa on silicon wafers.
\begin{table}
\caption{Refractive index of Ta$_2$O$_5$ films from different coaters.}
\begin{indented}
\lineup	
	\item[] \begin{tabular}{lcccc}
	\br
		& \multicolumn{2}{c}{as deposited}	& \multicolumn{2}{c}{annealed 500 $^\circ$C}\\
		& \multicolumn{1}{c}{1064 nm}		& \multicolumn{1}{c}{1550 nm}	& \multicolumn{1}{c}{1064 nm}	& \multicolumn{1}{c}{1550 nm}\\ 			\cline{2-5}
	GC		& 2.07 $\pm$ 0.01	& 2.06 $\pm$ 0.01	& 2.05 $\pm$ 0.01	& 2.03 $\pm$ 0.01\\
 	DIBS\footnote{Measured only with a spectrophotometer.}	& 2.06 $\pm$ 0.01	& 2.04 $\pm$ 0.01	& 2.03 $\pm$ 0.01	& 2.02 $\pm$ 0.01\\
 	SPECTOR	& 2.11 $\pm$ 0.01	& 2.09 $\pm$ 0.01	& 2.09 $\pm$ 0.01	& 2.07 $\pm$ 0.01\\
	\br
	\end{tabular}
\end{indented}
\end{table}
% ---
\begin{table}
\setlength{\tabcolsep}{3pt}
\caption{Mechanical parameters of Ta$_2$O$_5$ films from different coaters.}
\begin{indented}
\lineup	
	\item[] \begin{tabular}{lcccccc}
	\br 		
	 		&				& $\rho_c$ [g/cm$^3$]	& a [10$^{-4}$ rad Hz$^{-b}$]	& b					& $Y_c$ [GPa]	& $\nu_c$\\
	\mr
 	GC 		&				& 7.40 $\pm$ 0.03		& $4.61 \pm 0.11$				& 0.036 $\pm$ 0.003	& 121 $\pm$ 1	& 0.30 $\pm$ 0.01\\
 	DIBS	&				& 7.04 $\pm$ 0.09		& $8.20 \pm 0.24$				& -					& 117 $\pm$ 1	& 0.27 $\pm$ 0.01\\
 	SPECTOR	&				& 7.75 $\pm$ 0.03		& $7.60 \pm 0.21$				& 0.045 $\pm$ 0.003	& 121 $\pm$ 1	& 0.29 $\pm$ 0.01\\
 	\cline{1-2}
	GC		& 500 $^\circ$C	& 7.33 $\pm$ 0.06		& $1.88 \pm 0.06$				& 0.101 $\pm$ 0.004	& 117 $\pm$ 1	& 0.28 $\pm$ 0.01\\
	DIBS	& 500 $^\circ$C	& 6.94 $\pm$ 0.09		& $2.27 \pm 0.14$				& 0.078 $\pm$ 0.007	& 115 $\pm$ 1	& 0.28 $\pm$ 0.01\\
	SPECTOR	& 500 $^\circ$C	& 7.47 $\pm$ 0.09		& $2.29 \pm 0.06$				& 0.079 $\pm$ 0.003	& 121 $\pm$ 2	& 0.29 $\pm$ 0.01\\
	\br
	\end{tabular}
\end{indented}	
\end{table}
% ---
\begin{figure}
\centering
	\includegraphics{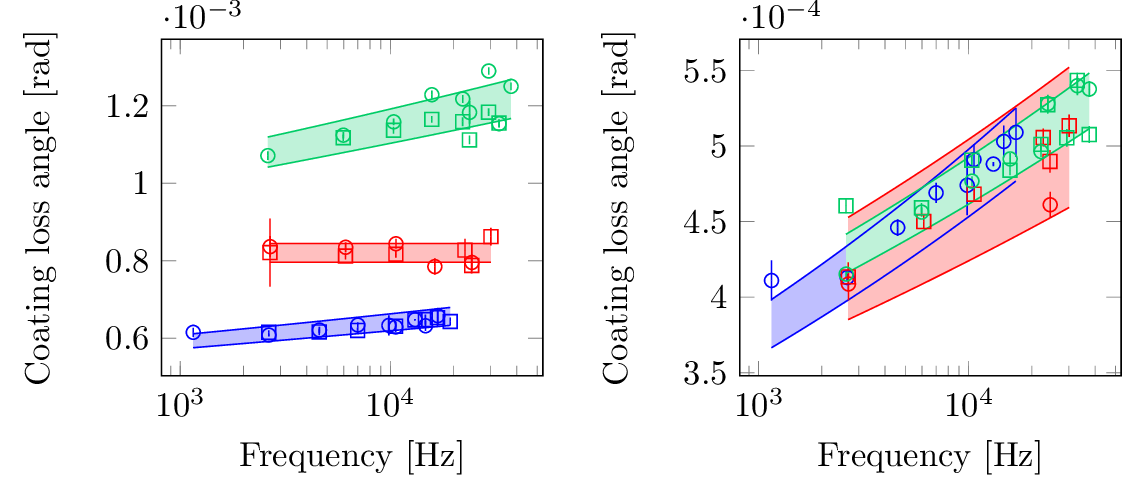}
	\caption{Mechanical loss of Ta$_2$O$_5$ films from different coaters: GC (blue), DIBS (red) and SPECTOR (green), before (left) and after the annealing (right). Different markers denote distinct samples; shaded regions represent uncertainties from fitting a frequency dependent loss model $\phi_c = af^{b}$ to each data set, via least-squares linear regression.}
\end{figure}
%
% -------------------------
%
\subsection{TiO$_2$}
Titania features a very high refractive index (n $\sim 2.3$), making it particularly well suited to increase the index contrast within the HR stack and thus consequently to allow for thinner high-index dissipative layers. Unfortunately, it becomes poly-crystalline when annealed at $T \geq 300$ $^\circ$C \cite{Lee06, Chen08}, yielding scattering and absorption losses far from meeting the stringent requirements of current gravitational-wave detectors \cite{{Pinard17}}. Because of crystallization, we could not measure its refractive index and some mechanical parameters after annealing.

Before annealing, the density of our film is slightly higher than that of typical sputtered titania films \cite{Bundesmann17} and close to that of the crystalline anatase phase (3.9 g/cm$^3$). The loss increases significantly after the annealing, very likely due to crystallization.
\begin{table}
\caption{Refractive index of the as-deposited TiO$_2$ film.}
\begin{indented}
\lineup	
	\item[] \begin{tabular}{lcc}
	\br
		& 1064 nm			& 1550 nm\\ 			\cline{2-3}
	GC	& 2.35 $\pm$ 0.05	& 2.33 $\pm$ 0.03\\
	\br
	\end{tabular}
\end{indented}
\end{table}
% ---
\begin{table}
\caption{Mechanical parameters of the TiO$_2$ film.}
\begin{indented}
\lineup	
	\item[] \begin{tabular}{lcccccc}
	\br 		
	 	&				& $\rho_c$ [g/cm$^3$]	& a [10$^{-4}$ rad Hz$^{-b}$]	& b					& $Y_c$ [GPa]	& $\nu_c$\\
	\mr
 	GC 	&				& 3.89 $\pm$ 0.06		& 2.05 $\pm$ 0.16				& 0.140 $\pm$ 0.008	& 145 $\pm$ 1	& 0.26 $\pm$ 0.01\\
 	\cline{1-2}
	GC	& 500 $^\circ$C	& -						& 0.46 $\pm$ 0.05				& 0.330 $\pm$ 0.011	& -				& -\\
	\br
	\end{tabular}
\end{indented}	
\end{table}
% ---
\begin{figure} 
\centering
	\includegraphics{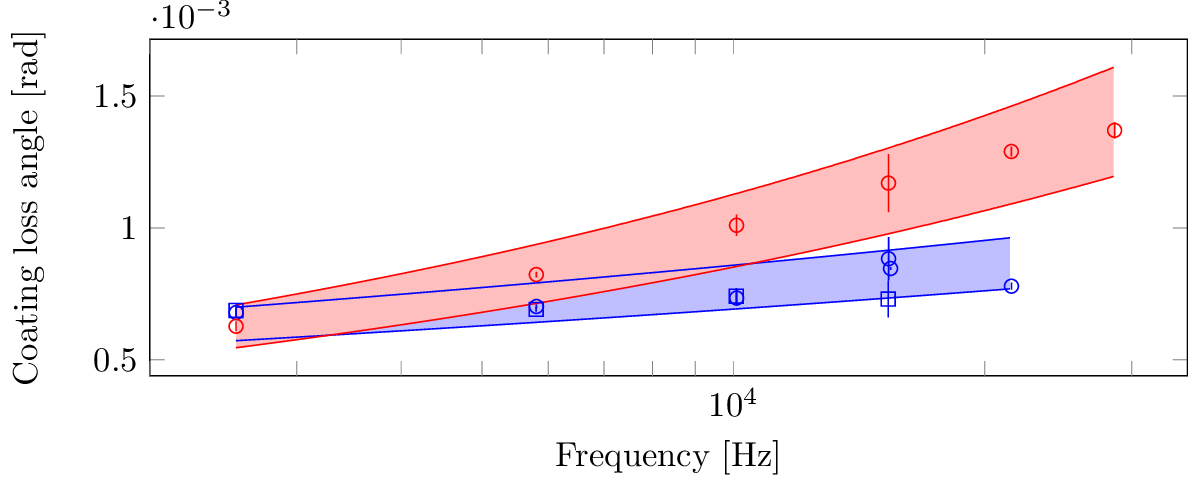}
	\caption{Mechanical loss of the TiO$_2$ film, before (blue) and after annealing (red). Different markers denote distinct samples; shaded regions represent uncertainties from fitting a frequency-dependent loss model $\phi_c = af^{b}$ to each data set, via least-squares linear regression.}
\end{figure}
%
% -------------------------
%
\subsection{Ta$_2$O$_5$-TiO$_2$ mixture}
The titania content in HR coatings for gravitational-wave detectors was initially determined by analyzing a set of Bragg reflectors with $\lambda/4$ layers \cite{Harry07}, produced in the DIBS and in the GC with different titania-to-tantala mixing ratios; Ti/Ta = 0.27 yielded minimum loss and had thus been chosen as the optimal ratio. Since then, the design of the HR coatings of Advanced LIGO and Advanced Virgo has evolved \cite{Pinard17, Granata16}, while the Ti/Ta ratio in the titania-doped tantala layers remained the same. This choice is now confirmed by our latest loss measurements of single titania-doped tantala films produced with the GC, which we also characterized through Rutherford back-scattering (RBS) and energy-dispersive X-ray (EDX) spectrometry.
\begin{figure} 
\centering
	\includegraphics{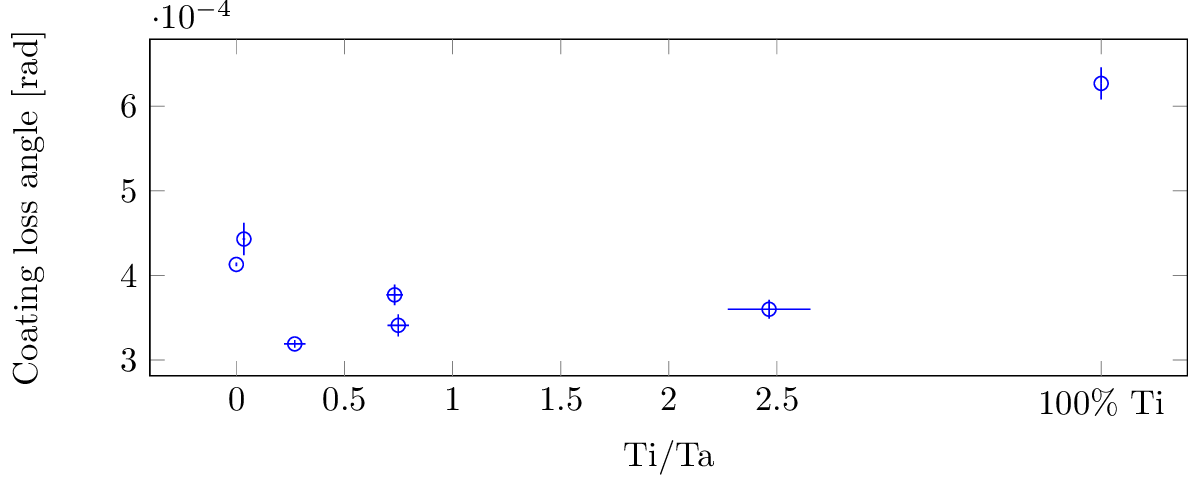}
	\caption{Mechanical loss of annealed Ta$_2$O$_5$-TiO$_2$ films deposited with the GC, as a function of the mixing ratio Ti/Ta; the data point of poly-crystalline TiO$_2$ is also shown, for comparison. For clarity, only values measured at $\sim2.5$ kHz are shown; the same trend has been observed at $\sim10$ kHz.}
\end{figure}

By comparing Ti/Ta = 0.27 titania-doped tantala to undoped GC tantala, we observe that both coating materials feature similar loss before annealing, whereas after annealing the loss of titania-doped tantala is $\sim25$\% lower in the whole sampled band, suggesting that the lower loss of titania-doped tantala is the result of a combined effect of mixing and annealing. The $\sim25$\% loss reduction has been confirmed by independent coating thermal noise measurements \cite{Gras18}. Except for the sample with Ti/Ta = 0.04, all the annealed titania-doped tantala coatings showed a lower loss than that of annealed undoped tantala; this is in fairly good agreement with molecular dynamics simulations showing that even small amounts of doping could decrease the coating internal friction \cite{Trinastic16}.

Several results of mechanical loss measurements are already available for titania-doped tantala coatings \cite{Harry06, Comtet07, Flaminio10, Villar10, Granata16, Principe15, Martin08, Martin09}, and a correlation between coating internal friction and microscopic structure has also been found \cite{Bassiri13}.

The fairest comparison with our results would be with annealed titania-doped tantala coatings deposited with the GC on cantilever blades \cite{Granata16} under identical conditions (deposition rate, Ti/Ta ratio), featuring a constant coating loss of \mbox{(2.4 $\pm$ 0.3)$\cdot 10^{-4}$ rad} in the \mbox{50-900 Hz} band; by extrapolating our results of the annealed samples to lower frequencies, we obtain \mbox{(2.2 $\pm$ 0.1)$\cdot 10^{-4}$ rad} at 50 Hz and \mbox{(3.0 $\pm$ 0.2)$\cdot 10^{-4}$ rad} at 900 Hz, matching the value from cantilever blades at 120 Hz. By correcting the cantilever blade results by our value of coating Young's modulus we would obtain 3.01 $\cdot$ 10$^{-4}$ rad, so that the match with our results would be shifted to 920 Hz.
\begin{table}
\caption{Refractive index of the Ta$_2$O$_5$-TiO$_2$ film with Ti/Ta = 0.27.}
\begin{indented}
\lineup	
	\item[] \begin{tabular}{lcccc}
	\br
		& \multicolumn{2}{c}{as deposited}	& \multicolumn{2}{c}{annealed 500 $^\circ$C}\\
		& \multicolumn{1}{c}{1064 nm}		& \multicolumn{1}{c}{1550 nm}	& \multicolumn{1}{c}{1064 nm}	& \multicolumn{1}{c}{1550 nm}\\ 			\cline{2-5}
	GC		& 2.11 $\pm$ 0.01	& 2.10 $\pm$ 0.01	& 2.09 $\pm$ 0.01	& 2.08 $\pm$ 0.01\\
	\br
	\end{tabular}
\end{indented}
\end{table}
% ---
\begin{table}
\caption{Mechanical parameters of the Ta$_2$O$_5$-TiO$_2$ film with Ti/Ta = 0.27.}
\begin{indented}
\lineup	
	\item[] \begin{tabular}{lcccccc}
	\br 		
	 	&				& $\rho_c$ [g/cm$^3$]	& a [10$^{-4}$ rad Hz$^{-b}$]	& b					& $Y_c$ [GPa]	& $\nu_c$\\
	\mr
 	GC 	&				& 6.87 $\pm$ 0.06		& 4.82 $\pm$ 0.18				& 0.029 $\pm$ 0.004	& 122 $\pm$ 1	& 0.30 $\pm$ 0.01\\
 	\cline{1-2}
	GC	& 500 $^\circ$C	& 6.65 $\pm$ 0.07		& 1.43 $\pm$ 0.07				& 0.109 $\pm$ 0.005	& 120 $\pm$ 4	& 0.29 $\pm$ 0.01\\
	\br
	\end{tabular}
\end{indented}	
\end{table}
% ---
\begin{figure} 
\centering
	\includegraphics{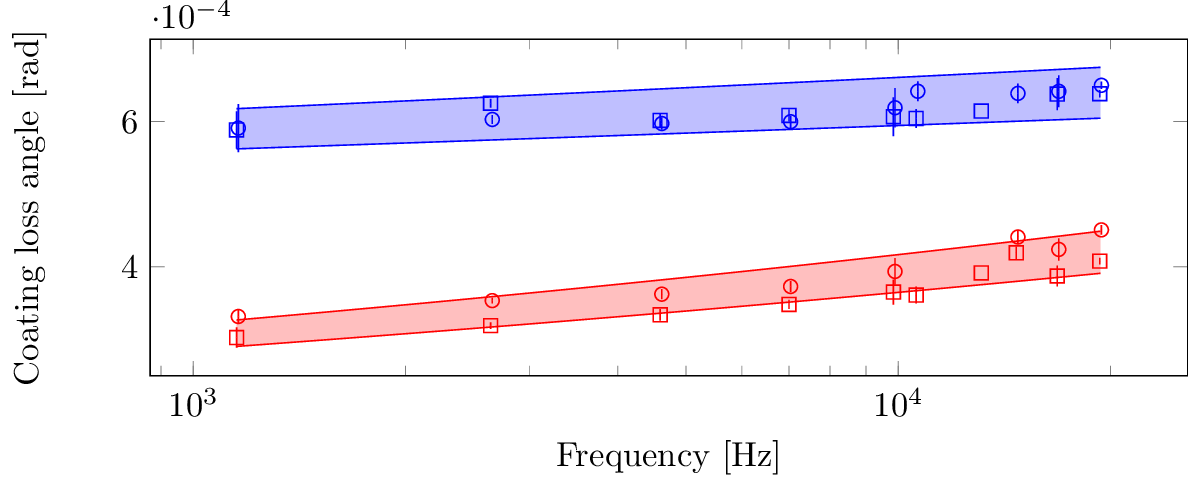}
	\caption{Mechanical loss of the Ta$_2$O$_5$-TiO$_2$ film with Ti/Ta = 0.27, before (blue) and after annealing (red). Different markers denote distinct samples; shaded regions represent uncertainties from fitting a frequency dependent loss model $\phi_c = af^{b}$ to each data set, via least-squares linear regression.}
\end{figure}

The refractive index of the Ta$_2$O$_5$-TiO$_2$ coating with Ti/Ta = 0.27 is in between those of tantala and titania, as expected \cite{Lee06}.  Its optical absorption is lower than that of tantala, as we could measure on single films as well as on the HR coatings of Advanced LIGO and Advanced Virgo \cite{Pinard17}; this in agreement with our previous measurements \cite{Flaminio10} and with previous studies of coatings deposited in similar conditions \cite{Lee06}, showing that the lowest absorption is obtained with low titania content.

The annealing increased the physical thickness of the coating, which in turn resulted in a decrease of density and of refractive index, as demonstrated by previous studies as well \cite{Lee06}. Finally, we observed that the crystallization temperature is the same of that of tantala, i.e. between 600 $^\circ$C and 650 $^\circ$C \cite{Amato18}, and hence much higher than that of titania \cite{Lee06, Chen08}.
%
% -------------------------
%
\subsection{SiO$_2$}
During the mechanical characterization of silica coatings, we observed a branching of the coating loss as a function of mode shape, closely following a separation in mode families. According to a model successfully tested on substrates \cite{Cagnoli18}, this phenomenon could be caused by spurious loss of the edge of the disks; this latter is also coated during deposition, since masking it would induce uncontrolled, undesired shadowing effects on the main  surfaces. According to this hypothesis, as the actual energy lost depends mainly on the vibration at the coated disk edge, the coating loss would be different for different mode shapes. Thus, given $ d \ll 1$ the thickness of a thin surface layer of the coated disk edge, and introducing the mode-dependent dilution factor density $\epsilon$ and the edge energy loss $\phi_e$, the frequency-dependent model for the coating loss of silica reads
\begin{equation}
\phi_c (f) = a f^b + \epsilon d\phi_e \ .
\end{equation}
If neglected, the edge effect could lead to a poor estimation of coating loss; this is particularly evident for the GC and DIBS samples, for which the $a f^b$ term is hidden by the overall measured trend. When considering the $a f^b$ term only, the loss of the GC and the DIBS samples appeared fairly constant ($b \sim 0$) before annealing, whereas the loss of the SPECTOR sample showed a weak decreasing trend ($b < 0$); after the annealing, the loss values of the SPECTOR and GC samples moved closer to those of the DIBS sample but kept their distinctive features. Indeed, once again, we observed that the SPECTOR sample had the fastest deposition rate (2 \AA/s) and the highest loss values; however, while having the same deposition rate (within 25\% experimental uncertainty) than the GC sample, the DIBS sample had the lowest loss values. This result might be related to the unusually small measured density value for the DIBS sample, and will be subject to further investigation.

Previous studies have been made of the loss \cite{Penn03, Comtet07, Villar10, Granata16, Principe15, Martin14, Hamdan14, Mariana19} and the elastic constants \cite{Cetinorgu09} of silica coatings. Recently we found a correlation between loss and microscopic structure, which holds for the silica coatings discussed here and for bulk fused silica as well \cite{Granata18}.

We may compare our results with those obtained from clamped cantilever blades \cite{Granata16} and free-standing micro-cantilevers made of coatings \cite{Mariana19}, all produced in the GC under identical conditions. The cantilever blades yielded a constant coating loss of \mbox{(4.5 $\pm$ 0.3)$\cdot 10^{-5}$ rad} in the \mbox{50-900 Hz} band; by considering the $a f^b$ term only and extrapolating the results of the annealed sample to lower frequencies, we obtain \mbox{(2.2 $\pm$ 0.7)$\cdot 10^{-5}$ rad} at 50 Hz and \mbox{(2.4 $\pm$ 1.0)$\cdot 10^{-5}$ rad} at 900 Hz, eventually matching the value from cantilever blades only at very high frequency. In other words, there is an irreconcilable discrepancy between our latest results obtained with disks and a GeNS system and our previous results obtained with clamped cantilever blades. It is likely that the loss measured with clamped cantilever blades was affected by spurious suspension losses (in the clamp and/or in the weld \cite{Granata16}), yielding to their systematic overestimation. The free-standing micro-cantilevers had been measured before annealing, yielding a variable loss of \mbox{(0.8 - 1.8)$\cdot 10^{-3}$ rad} in the \mbox{1.2-21.8 kHz} band \cite{Mariana19}; again, by considering the $a f^b$ term only and extrapolating the results of the annealed sample to the same band, we obtain \mbox{(1.6 $\pm$ 0.5)$\cdot 10^{-4}$ rad} at 1.2 kHz and \mbox{(1.7 $\pm$ 0.5)$\cdot 10^{-4}$ rad} at 21.8 kHz, in clear disagreement with the results from free-standing micro-cantilevers. Such a discrepancy can not be explained so far, and will be subject to further investigation.

Remarkably, the SPECTOR sample is significantly denser and stiffer than the GC one, whose properties closely resemble those of bulk fused silica \cite{McSkimin53}. 
\begin{table}
\caption{Refractive index of SiO$_2$ films from different coaters.}
\begin{indented}
\lineup	
	\item[] \begin{tabular}{lcccc}
	\br
		& \multicolumn{2}{c}{as deposited}	& \multicolumn{2}{c}{annealed 500 $^\circ$C}\\
		& \multicolumn{1}{c}{1064 nm}		& \multicolumn{1}{c}{1550 nm}	& \multicolumn{1}{c}{1064 nm}	& \multicolumn{1}{c}{1550 nm}\\ 			\cline{2-5}
	GC		& 1.47 $\pm$ 0.01	& 1.46 $\pm$ 0.01	& 1.45 $\pm$ 0.01	& 1.45 $\pm$ 0.01\\
	DIBS\footnote{Measured only with a spectrophotometer.}	& 1.44 $\pm$ 0.01 & 1.44 $\pm$ 0.02	& 1.44 $\pm$ 0.01	& 1.44 $\pm$ 0.02	\\
 	SPECTOR	& 1.48 $\pm$ 0.01	& 1.47 $\pm$ 0.01	& 1.47 $\pm$ 0.01	& 1.46 $\pm$ 0.01\\
	\br
	\end{tabular}
\end{indented}
\end{table}
% ---
\begin{table}
\caption{Mechanical parameters of SiO$_2$ films from different coaters.}
\begin{indented}
\lineup	
	\item[] \begin{tabular}{lcccc}
	\br 		
	 		&				& $\rho_c$ [g/cm$^3$]	& $Y_c$ [GPa]	& $\nu_c$\\
	\mr
 	GC 		&				& 2.33 $\pm$ 0.06		& 66 $\pm$ 4	& 0.19 $\pm$ 0.02\\
 	DIBS	&				& 2.02 $\pm$ 0.09		& 74 $\pm$ 2	& 0.18 $\pm$ 0.02\\
 	SPECTOR	&				& 2.38 $\pm$ 0.01		& 78 $\pm$ 1	& 0.14 $\pm$ 0.01\\
 	\cline{1-2}
	GC		& 500 $^\circ$C	& 2.20 $\pm$ 0.04		& 70 $\pm$ 1	& 0.19 $\pm$ 0.01\\
	DIBS	& 500 $^\circ$C	& 1.91 $\pm$ 0.09		& 75 $\pm$ 2	& 0.19 $\pm$ 0.02\\
	SPECTOR	& 500 $^\circ$C	& 2.36 $\pm$ 0.03		& 78 $\pm$ 1	& 0.11 $\pm$ 0.01\\
	\br
	\end{tabular}
\end{indented}	
\end{table}
% ---
\begin{table}
\caption{Mechanical loss of SiO$_2$ films from different coaters.}
\begin{indented}
\lineup	
	\item[] \begin{tabular}{lcccc}
	\br 		
	 		&				& a [10$^{-4}$ rad Hz$^{-b}$]	& b						&  $d\phi_e$ [10$^{-6}$ m]\\
	\mr
 	GC 		&				& 1.37 $\pm$ 0.22				& 0.024 $\pm$ 0.019		& 3.11 $\pm$ 0.26\\
 	DIBS	&				& 1.45 $\pm$ 0.06				& -0.016 $\pm$ 0.005	& 0.39 $\pm$ 0.03\\
 	SPECTOR	&				& 8.87 $\pm$ 0.19				& -0.083 $\pm$ 0.003	& 0.84 $\pm$ 0.11\\
 	\cline{1-2}
	GC		& 500 $^\circ$C	& 0.20 $\pm$ 0.04				& 0.030 $\pm$ 0.024		& 1.41 $\pm$ 0.05\\
	DIBS 	& 500 $^\circ$C	& 1.42 $\pm$ 0.54				& -0.208 $\pm$ 0.044	& 0.42 $\pm$ 0.05\\
	SPECTOR	& 500 $^\circ$C	& 1.26 $\pm$ 0.23				& -0.069 $\pm$ 0.024	& 0.19 $\pm$ 0.15\\
	\br
	\end{tabular}
\end{indented}	
\end{table}
% ---
\begin{figure} 
\centering
	\includegraphics{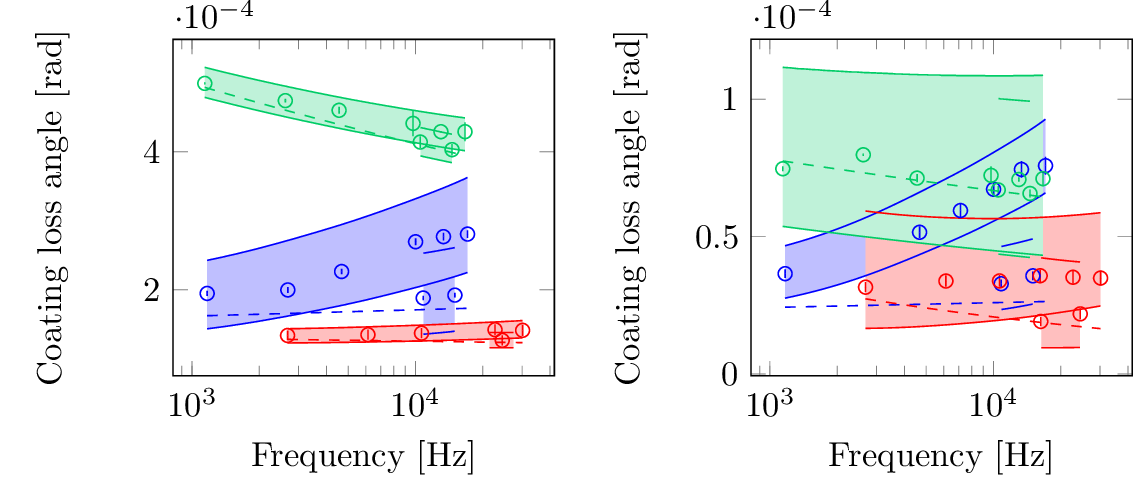}
	\caption{Mechanical loss of SiO$_2$ films from different coaters: GC (blue), DIBS (red) and SPECTOR (green), before (left) and after the annealing (right); shaded regions represent uncertainties from fitting a frequency-dependent loss model $\phi_c = af^{b} + \epsilon d\phi_e$ to each data set, via numerical non-linear regression; dashed curves show the behavior of the $af^{b}$ term only.}
\end{figure}
%
% -------------------------
%
\subsection{HR coatings of Advanced LIGO and Advanced Virgo}
The reference loss values of the Advanced LIGO and Advanced Virgo input (ITM) and end (ETM) mirror HR coatings had been previously estimated \cite{Granata16} by assuming $Y_c = 140$ GPa for titania-doped tantala layers \cite{Abernathy14}. These values may now be updated by using our measured value of Young's modulus of titania-doped tantala layers, $Y_c = 120$ GPa: the new estimations, listed in Table \ref{TABLElossHRstacks}, are about 10\% higher than the previous ones \cite{Granata16} for both ITM and ETM coatings. 
\begin{table}
\caption{Nominal specifications of Advanced LIGO and Advanced Virgo input (ITM) and end (ETM) mirror HR coatings: transmission $T$, number of layers $N$, thickness of titania-doped tanatala layers $t_H$, thickness of silica layers $t_L$, thickness ratio $r = t_H/t_L$, total thickness $t = t_H + t_L$.}
\begin{indented}
\lineup	
	\item[] \begin{tabular}{lcccccc}
	\br 		
			& $T$	& $N$	& $t_H$ [nm]	& $t_L$ [nm]	& $r$	& $t$ [nm]\\
	\mr
		ITM	& 1.4\%	& 18	& 727			& 2080			& 0.32	& 2807\\
 		ETM	& 4 ppm	& 38	& 2109			& 3766			& 0.56	& 5875\\
	\br
	\end{tabular}
\end{indented}	
\end{table}
% ---
\begin{table}
\caption{\label{TABLElossHRstacks}Mechanical loss of Advanced LIGO and Advanced Virgo input (ITM) and end (ETM) mirror HR coatings. $(r, a)_k$ is the pair denoting the $k$-th mode with $r$ radial and $a$ azimuthal nodes, $\phi^c_k$ is the coating loss of the $k$-th mode defined in Eq.(\ref{EQcoatLossKth}).}
\begin{indented}
\lineup	
	\item[] \begin{tabular}{lclccc}
	\br 		
			& $f [HZ]$		& $(r, a)_k$	& $\phi^c_k$ [10$^{-4}$ rad]	& a [10$^{-4}$ rad Hz$^{-b}$]	& b\\
	\mr
		ITM	& 2708.1	& (0, 2)$_1$	& 1.6 $\pm$ 0.1					& 1.1 $\pm$ 0.3			& 0.05 $\pm$ 0.03\\
			& 16092.6	& (0, 5)$_{10}$	& 1.7 $\pm$ 0.1					&	&\\
			& 16283.9	& (1, 2)$_{12}$	& 1.8 $\pm$ 0.1					&	&\\
			& 22423.4	& (0, 6)$_{15}$	& 1.7 $\pm$ 0.1					&	&\\
 	\cline{1-2}
 		ETM	& 2708.6	& (0, 2)$_1$	& 2.4 $\pm$ 0.1					& 2.2 $\pm$ 0.6			& 0.01 $\pm$ 0.03\\
 			& 6168.3	& (0, 3)$_4$	& 2.3 $\pm$ 0.1					&	&\\
 			& 16088.1	& (0, 5)$_{10}$	& 2.5 $\pm$ 0.1					&	&\\
 			& 16297.9	& (1, 2)$_{12}$	& 2.4 $\pm$ 0.1					&	&\\
 			& 22414.5	& (0, 6)$_{15}$	& 2.3 $\pm$ 0.1					&	&\\
	\br
	\end{tabular}
\end{indented}	
\end{table}

\begin{figure} 
\centering
	\includegraphics{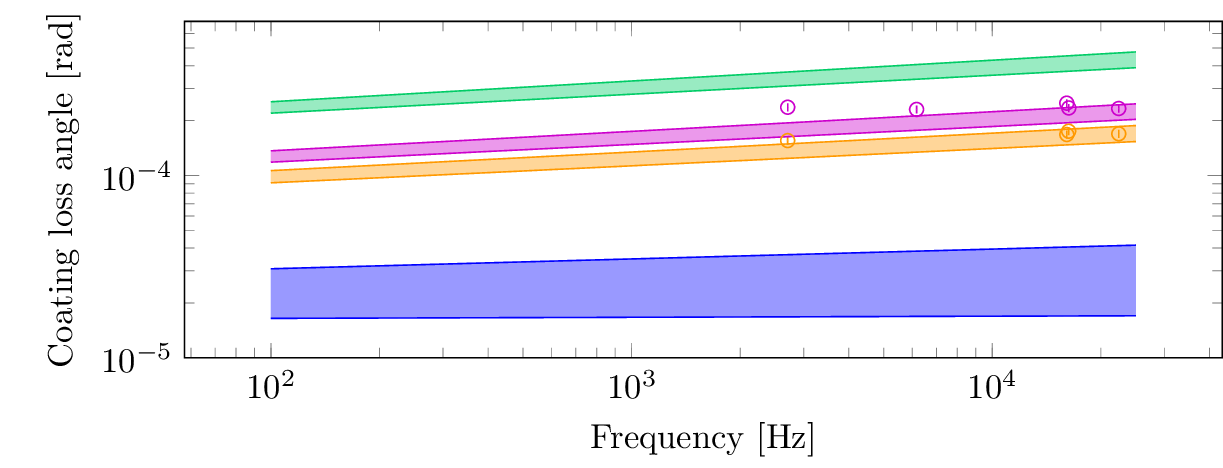}
	\caption{Mechanical loss of Advanced LIGO and Advanced Virgo input (ITM, yellow) and end (ETM, purple) mirror HR coatings: comparison between the expected values (shaded regions) calculated via Eq.(\ref{EQexpectHR}) and measured values (markers) from Table \ref{TABLElossHRstacks}. The frequency-dependent loss term $\phi_c(f) = af^{b}$ of Ta$_2$O$_5$-TiO$_2$ (green) and SiO$_2$ (blue) layers, from Table \ref{TABLEmechParGC}, is also shown for comparison.}
	\label{FIGstacksHR}
\end{figure}

It is commonly accepted that the loss of a HR coating stack $\phi_{HR}$ is the linear combination of the measured loss of its constituent layers \cite{Penn03},
\begin{equation}
\label{EQexpectHR}
\phi_{\textrm{\tiny{HR}}} = \frac{t_{\textrm{\tiny{H}}} Y_{\textrm{\tiny{H}}} \phi_{\textrm{\tiny{H}}} + t_{\textrm{\tiny{L}}} Y_{\textrm{\tiny{L}}} \phi_{\textrm{\tiny{L}}}}{t_{\textrm{\tiny{H}}} Y_{\textrm{\tiny{H}}} + t_{\textrm{\tiny{L}}} Y_{\textrm{\tiny{L}}}}\ ,
\end{equation}
where $t_{\textrm{\tiny{H}}}$, $Y_{\textrm{\tiny{H}}}$ and $\phi_{\textrm{\tiny{H}}}$ ($t_{\textrm{\tiny{L}}}$, $Y_{\textrm{\tiny{L}}}$ and $\phi_{\textrm{\tiny{L}}}$) are the thickness, the Young's modulus and the loss of the high-index Ta$_2$O$_5$-TiO$_2$ (low-index SiO$_2$) layers. Fig. \ref{FIGstacksHR} shows the comparison between the expected loss of Eq.(\ref{EQexpectHR}), calculated using the values of loss and Young's modulus of Table \ref{TABLEmechParGC}, and the updated loss of Table \ref{TABLElossHRstacks}, obtained from direct loss measurements of the HR coatings \cite{Granata16}. The measured loss is fairly constant over the sampled band (2.7-22.4 kHz), whereas the expectations have the same frequency dependence of their dominant contribution, the titania-doped tantala layers. When extrapolating down to 0.1 kHz, i.e. the region of the Advanced LIGO and Advanced Virgo detection band limited by coating thermal noise, the expectations  underestimate the actual measured loss of the HR coatings, by about 30\% for the ITM coating and about 43\% for the ETM coating. Though well-known \cite{Granata16}, this discrepancy remains unexplained to date and will be subject to further investigation.
%
% ---------------------------------------------
%
\section{Conclusions}
We presented in this work the results of an extensive campaign of optical and mechanical characterization of the IBS oxide layers (Ta$_2$O$_5$, TiO$_2$, Ta$_2$O$_5$-TiO$_2$, SiO$_2$) within the HR coatings of the Advanced LIGO, Advanced Virgo and KAGRA gravitational-wave detectors. These layers, deposited at the LMA, have been used for the first observing runs (designated O1 and O2) of Advanced LIGO and Advanced Virgo \cite{Abbott18}, when the first detections of mergers of black-hole and neutron-star binaries occurred.

Our measurements provide several coating parameters which are required to predict the power spectral density of coating thermal noise in gravitational-wave interferometers: refractive index, thickness, optical absorption, composition, density, internal friction and elastic constants have been measured; the mechanical parameters have been measured with a GeNS system \cite{Cesarini09}, which has now been adopted as a standard apparatus by the Virgo and LIGO Collaborations \cite{Granata16, Vajente17}. The main outcomes may be summarized as follows:
\begin{description}
\item [Frequency-dependent loss] Measurements of coating internal friction performed with our GeNS system \cite{Amato18, GranataLVC1603} showed weak but clear frequency-dependent trends, i.e. $\phi_c(f) = af^{b}$ with $-0.208 < b < 0.140$, depending on the sample considered; the frequency dependence observed for Ta$_2$O$_5$ and Ta$_2$O$_5$-TiO$_2$ layers has been later confirmed by independent measurements of coating thermal noise on the end mirror (ETM) HR coatings of Advanced LIGO and Advanced Virgo \cite{Gras18}. Our method for loss characterization is exclusively based on measured quantities (quality factors, frequencies and masses) and, unlike other experimental setups based on the ring-down method, does not require prior knowledge of the coating Young's modulus and thickness.
\item [Deposition rate] With our tantala films, we observed that the level of coating internal friction is determined by the deposition rate: the slower the rate, the lower the loss; the same rule fairly holds for our silica films too. Thus, special care should be taken when comparing the loss of coating samples, even if composed of the same coating material.
\item [Annealing] Post-deposition heat treatment makes the thickness increase and hence density decrease, while optical absorption, refractive index and internal friction decrease. Depending on the initial values considered, the reduction of internal friction is of a factor 1.5 to 2.5 for Ta$_2$O$_5$ films, 1.8 for the Ta$_2$O$_5$-TiO$_2$ layer with Ti/Ta = 0.27, 5 to 6.5 for SiO$_2$ films. As the disks are annealed at 900 $^\circ$C before deposition, the observed loss change upon 500 $^\circ$C annealing is due to the coating only. Once annealed, all the films Ta$_2$O$_5$ exhibited equal loss, as if their deposition history had been erased, whereas the gap between the loss values of SiO$_2$ films decreased. For SiO$_2$ layers, it has been possible to establish a correlation between their structural change upon annealing and their internal friction \cite{Granata18}; for Ta$_2$O$_5$ and Ta$_2$O$_5$-TiO$_2$ films, we carried out analogous studies of structure and internal friction as a function of annealing temperature and duration \cite{Amato18, CoilletPhD} but --despite the large variation of internal friction-- we observed very limited structural change and eventually found no correlations.
\item [Elastic constants] Our GeNS system allows the estimation of coating Young's modulus and Poisson's ratio, via the measurement of dilution factors \cite{Granata15}. Our values of Young's modulus for  Ta$_2$O$_5$ and Ta$_2$O$_5$-TiO$_2$ coatings are 14\% lower than those commonly accepted \cite{Cetinorgu09, Abernathy14} by the scientific collaborations running gravitational-wave interferometers. Two reasons may explain this discrepancy: likely the different nature of coating samples used in previous works, determined by the different deposition parameters used, and the method used, i.e. nano-indentations. The analysis of nano-indentation measurements relies on the prior knowledge of the coating Poisson's ratio and is model dependent. Moreover, nano-indentations of our tantala SPECTOR coatings suggest that the outcome of this technique may depend on the nature of the substrate: we obtained a reduced Young's modulus of 100 GPa from the coating sample deposited on a fused silica witness substrate, of 130 GPa from the coating sample deposited on a silicon wafer; both samples were from the same deposition run. While the discussion of the details about those measurements and their analysis is beyond the scope of this article, the important point we would like to stress here is that the outcomes of nano-indentations should be considered cautiously. Unlike nano-indentations, our estimation method is a non-destructive technique which does not require prior knowledge of the coating Poisson's ratio; it has been tested also on tantala SPECTOR coatings deposited at the same time on fused-silica disks and on silicon wafers, yielding consistent results. The agreement between finite-element simulations and measured dilution factors vouches for the reliability of our method; however, larger residuals for silica coatings seem to point out that either our model or our simulations might need further fine tuning, and will be be subject to further investigation.
\item [Edge effect] SiO$_2$ films showed a mode-dependent loss branching, which may be accounted for by including a term of spurious loss from the coated disks' edge in the coating loss model, as it is already the case for bare substrates \cite{Cagnoli18}.
\item [Loss of HR coatings] The reference loss values of the Advanced LIGO and Advanced Virgo input (ITM) and end (ETM) mirror HR coatings \cite{Granata16} have been updated by using our estimated value of Young's modulus of titania-doped tantala layers ($Y_c = 120$ GPa) and are about 10\% higher than previous estimations. By using our latest measurements of SiO$_2$ and  Ta$_2$O$_5$-TiO$_2$ films deposited on fused silica disks and the updated loss values of the Advanced LIGO and Advanced Virgo HR coatings, we demonstrated that the loss of a HR coating stack is not equal to the linear combination of the measured loss of its constituent layers: a different frequency dependence makes the linear combination smaller than the measured loss at lower frequencies (below 2 kHz). This is a confirmation of what had already been observed with the same coatings (deposited with the same conditions) measured on clamped cantilever blades \cite{Granata16}.
\item [Metrology issue] The frequency dependence of coating loss of Ta$_2$O$_5$, Ta$_2$O$_5$-TiO$_2$ and SiO$_2$ coatings had not been observed previously with clamped cantilever blades \cite{Comtet07, Flaminio10, Granata16}. Though our latest loss measurements of Ta$_2$O$_5$ and Ta$_2$O$_5$-TiO$_2$ layers are compatible with previous estimations at least in some specific frequency band, in the case of SiO$_2$ coatings there is a discrepancy that cannot be accounted for to date and which will be subject to further investigation.
\end{description}

These current coating layers are now being used for the ongoing joint observing run (O3) of Advanced LIGO and Advanced Virgo \cite{Abbott18} with the possible participation of KAGRA, which started on April 1$^{\textrm{{\tiny{st}}}}$ 2019. They will be used also for the following run (O4), in which KAGRA will fully join the present network of detectors.

Coating thermal noise is expected to be a severe limitation for present \cite{aLIGO, AdVirgo} and future \cite{Hild11, Abernathy11} ground-based gravitational-wave interferometers. This is the reason why, in the context of a medium term plan to improve the design sensitivity of their detectors (the projects are called {\it A+} and {\it Advanced Virgo +}), the LIGO and Virgo Scientific Collaborations  have already planned to produce new sets of mirrors with lower coating thermal noise, to be used for the next observing run (O5).  

More generally, coating thermal noise is already a fundamental limit for a large number of precision experiments based on optical and quantum transducers, such as opto-mechanical resonators \cite{Aspelmeyer14}, frequency standards \cite{Matei17} and quantum computers \cite{Martinis05}. Thus, in the context of a world-wide research effort devoted to solve this issue, the development of low-thermal-noise amorphous optical coatings is a long-term activity of the LMA.

Besides lowering the temperature of the mirrors \cite{KAGRA, Hild11, Abernathy11, Abbott17}, there are three key properties that may reduce coating thermal noise \cite{Harry02}: coating thickness and internal friction, which depend on intrinsic properties of coating materials, and laser beam size, which requires the development of deposition technology and larger substrates. Coating thickness is in turn a monotonically decreasing function of the refractive index contrast $c = n_H - n_L$ in the HR stack, where $n_H$ and $n_L$ are the high and low refractive indices, respectively; thus the larger $c$, the lower the coating thickness and hence the coating thermal noise (at constant reflection). As a consequence, the optimal coating materials would feature the lowest internal friction and the largest index contrast at the same time, and, in order to limit thermal lens effects, an optical absorption at least as low as it is to date. By taking the results from Ta$_2$O$_5$-TiO$_2$ and SiO$_2$ coatings presented here as a reference (see Tables \ref{TABLEoptParGC} and \ref{TABLEmechParGC}), future low-thermal-noise coating layers should have $c > 0.64$, $\phi_c < 2.4 \cdot 10^{-4}$ at $10^2$ Hz and $k \sim 10^{-7}$ at $\lambda_0=1064$ nm or $\lambda=1550$ nm.
%
% ---------------------------------------------
%
\section{Acknowledgments}
The authors would like to thank C. Bernard, S. Gavarini and N. Millard-Pinard of the {\it Institut de Physique des 2 Infinis de Lyon} for the RBS measurements and E. Coillet for the data analysis; R. Brescia, A. Scarpellini and M. Prato of the {\it Istituto Italiano di Tecnologia} for the EDX and XPS analyses; M. Neri of the {\it OPTMATLAB} for the preliminary ellipsometric analysis; S. Pavan and J.-L. Loubet of the {\it Laboratoire de Tribologie et Dynamique des Syst\`emes} for the nano-indentation analysis.
%
% ---------------------------------------------
%

\end{document}